\documentclass[12pt]{article}
\usepackage{amsfonts,amsmath,appendix,color,amssymb,graphicx}
\DeclareGraphicsExtensions{.pdf}

\makeatletter
\@addtoreset{equation}{section}
\makeatother
\newtheorem{theorem}{Theorem}[section]

\newtheorem{remark}{Remark}[section]
\newtheorem{proposition}{Proposition}[section]

\newtheorem{lemma}{Lemma}[section]
\newtheorem{corollary}{Corollary}[section]
\newtheorem{definition}{Definition}[section]
\def\br{\begin{remark}\rm\small}
\def\er{\end{remark}}
\def\bt{\begin{theorem}}
\def\et{\end{theorem}}
\def\bd{\begin{definition}}
\def\ed{\end{definition}}
\def\bp{\begin{proposition}}
\def\ep{\end{proposition}}
\def\bl{\begin{lemma}}
\def\el{\end{lemma}}
\def\bc{\begin{corollary}}
\def\ec{\end{corollary}}
\def\beaq{\begin{eqnarray}}
\def\eeaq{\end{eqnarray}}

\newcommand{\nn}{\nonumber}
\newcommand{\dd}{\mathrm{d}}
\newcommand{\beq}{\begin{equation}}
\newcommand{\eeq}{\end{equation}}
\newcommand{\bea}{\begin{eqnarray}}
\newcommand{\eea}{\end{eqnarray}}

\definecolor{rouge}{rgb}{0.84,0.18,0.07}
\definecolor{bleu}{rgb}{0.22,0.41,0.74}
\definecolor{vertf}{rgb}{0.08,0.46,0.07}
\usepackage[pdftex]{hyperref}
\hypersetup{colorlinks,urlcolor=vertf,citecolor=rouge,linkcolor=bleu,filecolor=black}

\begin{document}

\sloppy

\pagestyle{empty}

\addtolength{\baselineskip}{0.20\baselineskip}
\begin{center}
\vspace{26pt}
{\large \bf {Right tail asymptotic expansion \\ of Tracy-Widom beta laws}}
\end{center}

\vspace{26pt}

\begin{center}
{\sl Ga\"etan Borot}\footnote{Department of Mathematics, University of Geneva, \href{mailto:gaetan.borot@unige.ch}{\textsf{gaetan.borot@unige.ch}}}, 
{\sl C\'eline Nadal}\footnote{Rudolf Peierls Centre for Theoretical Physics, University of Oxford, and All Souls College (Oxford),
\href{mailto:celine.nadal@physics.ox.ac.uk}{\textsf{celine.nadal@physics.ox.ac.uk}}}
\end{center}

\vspace{20pt}
\begin{center}
{\bf Abstract}
\end{center}

\textsf{Using loop equations, we compute the large deviation function of the maximum eigenvalue to the right of the spectrum in the Gaussian $\beta$ matrix ensembles, to all orders in $N$. We then give a physical derivation of the all order asymptotic expansion of the right tail Tracy-Widom $\beta$ laws, for all $\beta > 0$, by studying the double scaling limit.}

%

\vspace{0.5cm}

\vspace{0.5cm}

\newpage


\vspace{26pt}
\pagestyle{plain}
\setcounter{page}{1}
\addtocounter{footnote}{-2}


\section{Introduction}
\label{sec:intro}

Extreme value statistics is an important issue in probability theory, 
which has a lot of applications in various fields,
such as geology (earthquakes, etc), meteorology (extreme floods, etc) 
and hydrology, the question of establishing safety norms for construction works,   growth problems in physics, finance, etc. 
Fischer and Tippett \cite{FT28}, completed by Gnedenko \cite{Gne43}, have shown that the distribution of the
maximum of a large number of independent identically distributed random variables falls in one of the three universality classes 
known as Weibull, Fr\'echet and Gumbel, depending on the law of the initial variables
(cf. also the book by Gumbel \cite{Gum58}).
 However, the classification of limit laws for the maximum of a large number of correlated variables is a hard and still open problem. A possible strategy is to deal with particular models of correlated random variables, determine the limit law for the maximum, and try to show its universality within the model.
 
 The eigenvalues of random matrices provide a good example of such highly correlated random variables.
 For the Gaussian ensembles, ie for $N\times N$ matrices with independent Gaussian entries,
 the distribution of the maximal eigenvalue is known to converge in the limit $N\to\infty$ 
 to the famous Tracy-Widom laws. Since the seminal work by Tracy and Widom \cite{TW92,TW95},
 these laws have appeared
 in very different and surprising contexts,
 ranging from statistical physics and probability theory all the way to
growth models \cite{Sph,Takeuchi} and biological sequence matching problems (for reviews see~\cite{AD,TracyReview,Deiftuni,maj-review,forrester,krug} and references thereof).



For the usual Gaussian matrix ensembles GOE (Gaussian Orthogonal Ensemble), GUE (Gaussian Unitary Ensemble)
and GSE (Gaussian Symplectic Ensemble), the limiting law of the maximal eigenvalue is given by the Tracy-Widom laws
with Dyson index $\beta=1,2$ or $4$ (resp. for the GOE, GUE or GSE), which
have been studied a lot.
One can also consider more generally Gaussian beta ensembles for $\beta >0$ (G$\beta$E).
In this case also it has been shown that the maximal eigenvalue converges to  a beta Tracy-Widom law $\mathsf{TW}_{\beta}(s)$,
but very few is known about these general laws.
The goal of our paper is to study in details the asymptotics of these laws for $s\to \infty$ (right tail).

For that purpose we actually compute first the large deviations of the maximal eigenvalue in the G$\beta$E,
which describe its large fluctuations on the right of the mean value (rare events).
We expect that the large deviation matches smoothly the Tracy-Widom law,
which describes the small typical fluctuations around the mean value.
By zooming close to the mean value in the expression of the right large deviation, we thus get
the right asymptotics of the Tracy-Widom laws for $\beta >0$.

In the first section, we review some known results about the Tracy-Widom beta laws, for $\beta=1,2,4$
and then for general $\beta >0$.
In the second section, we explain our method and give our results about the right large deviation
of the maximal eigenvalue in the $G\beta E$ and the right tail of the Tracy-Widom beta laws.


\subsection{Usual Tracy-Widom laws}

In random matrix theory, the Gaussian ensembles are one of the first proposed and most studied ensembles.
A Gaussian random matrix is a $N \times N$ matrix $\Phi$, which belongs to some vector space $E_{\beta}$, and whose entries are independent, identically distributed Gaussian variables. The distribution is proportional to $\dd \Phi\,e^{-\frac{N\beta}{2}\,\mathrm{Tr}\,V(\Phi)}$ with $V(\Phi) = \frac{\Phi^2}{2}$, and is invariant by conjugation under a group $G_{\beta}$. $\beta \in \{1,2,4\}$ is called the Dyson index: $\beta = 1$ defines the ensemble named GOE, for which $E_{1}$ is the space of real symmetric matrices and $G_1$ is the orthogonal group ; $\beta = 2$ defines the GUE, for which $E_2$ is the space of hermitian matrices and $G_2$ is the unitary group ; $\beta = 4$ defines the GSE, for which $E_4$ is the space of quaternionic self-dual matrices and $G_4$ the unitary symplectic group \cite{Dyson,MehtaBook}. The distribution induced on eigenvalues is then given by:
\beq\label{mes}
\dd\mu_{\beta,N}^{V}(\lambda_1,\ldots,\lambda_N) = \frac{1}{Z_{N,\beta}^{V}}\,\prod_{i = 1}^N \dd\lambda_i\,e^{-\frac{N\beta}{2}\,V(\lambda_i)}\,\prod_{1 \leq i < j \leq N} |\lambda_i - \lambda_j|^{\beta}
\eeq
Because of the Vandermonde determinant $\Delta(\lambda_i)=\prod_{i<j}(\lambda_j - \lambda_i)$ appearing to a power $\beta$ in the distribution \eqref{mes} for $G\beta E$, the eigenvalues are strongly correlated random variables. The distribution of the maximal eigenvalue when $N$ is large is thus of special interest, and is very different from the possible limit distribution for the maximum of a large number of independent random variables, which are Gumbel, Fr\'echet or Weibull distributions.

The fluctuations of the maximum eigenvalue in $\mathrm{G}\beta\mathrm{E}$ follow when $N$ is large, upon rescaling, the celebrated Tracy-Widom laws. Thanks to orthogonal polynomials and integrable properties, these laws are very well understood when $\beta = 1,2$ or $4$. Tracy and Widom have indeed shown more than 15 years ago the following theorem:
\begin{theorem} \cite{TW92,TW95}
\label{thm1}In the G$\beta$E for $\beta  \in \{1,2,4\}$, $\frac{\lambda_{\mathrm{max}} - 2}{c_{\beta}\,N^{2/3}}$ converges in law to a random variable $S$ distributed as:
\beq
\mathbb{E}(S \leq s) = \mathsf{F}_{\beta}(s)
\eeq
Let $q(s)$ be the Hastings-McLeod solution of Painlev\'e II \cite{HMcL}:
\beq
q''(s) = 2q^3(s) + sq(s),\qquad q(s) \mathop{\sim}_{s \rightarrow \infty} \mathrm{Ai}(s) \sim \frac{s^{-1/4}}{2\sqrt{\pi}}\,\exp\Big(-\frac{2s^{3/2}}{3}\Big) 
\eeq
Define $R(s) = \int_{s}^{\infty} q^2(t)\,\dd t = \big(q'(s)\big)^2 - sq^2(s) - q^4(s)$ and:
\beq
\mathsf{E}(s) = \exp\Big(-\frac{1}{2}\int_{s}^{\infty} q(t)\,\dd t\Big),\qquad \mathsf{H}(s) = \exp\Big(-\frac{1}{2}\int_{s}^{\infty} R(t)\,\dd t\Big)
\eeq
Then:
\bea
c_1 = 1 & & \mathsf{F}_1(s) = \mathsf{E}(s)\,\mathsf{H}(s) \\
c_2 = 1 & & \mathsf{F}_2(s) = \mathsf{H}^2(s) \\
c_4  = 2^{-2/3} & & \mathsf{F}_4(s) = \frac{1}{2}\big(\mathsf{E}(s) + \mathsf{E}^{-1}(s)\big)\mathsf{H}(s) 
\eea
\end{theorem}
Furthermore, Deift and Gioev \cite{DG07} have proved that, for polynomial $V$ and when the large $N$ spectral density behaves as a squareroot at the edge of the spectrum (this is the generic behavior), the fluctuations of $\lambda_{\mathrm{max}}$ are still of order $N^{-2/3}$ and follow (up to rescaling) the Tracy-Widom laws. Actually, they establish universality at the edge of the spectrum for the correlation densities of $n$ eigenvalues, for any $n \geq 1$. Those statistics can thus be expressed as in the G$\beta$E, in terms of the Airy kernel \cite{MehtaBook}.

\subsection{$\beta$ Tracy-Widom laws}

The expressions in Thm.~\ref{thm1} make clear that the distribution of the rescaled fluctuations of $\lambda_{\mathrm{max}}$ is very sensible to the value of $\beta$, and one may wish to study the probability density in Eqn.~\ref{mes} for arbitrary $\beta > 0$. Dumitriu and Edelman \cite{DE02} have shown that Eqn.~\ref{mes} with arbitrary $\beta$ can be realized by eigenvalues of a random tridiagonal matrix, whose entries are independent but accurately chosen random variables. Given the lack of known integrability properties when $\beta \neq \{1,2,4\}$, this connection with random tridiagonal matrices and thus to a diffusion process has been the main source of results. 

Rider, Ram\'{i}rez and Vir\'{a}g have shown the existence of a Tracy-Widom $\beta$ law with the same rescaling:
\begin{theorem}\cite{RRV}
In the G$\beta$E for any $\beta > 0$, $\frac{\lambda_{\mathrm{max}} - 2}{\,N^{2/3}}$ converges in law towards a random variable $S$. The Tracy-Widom $\beta$ law can be defined as its repartition function:
\beq
\mathbb{E}[S \leq s] = \mathsf{TW}_{\beta}(s)
\eeq
\end{theorem}
Notice that the definition of $\mathsf{TW}_{\beta}$ differs from the definition of $\textsf{F}_{\beta}(s)$ when $\beta = 4$ by a rescaling of the variable $s$, i.e. $\mathsf{TW}_{4}(s) = \mathsf{F}_4(2^{2/3}s)$. Later, Vir\'{a}g and Bloemendal have provided a characterization of $\mathsf{TW}_{\beta}$:
\begin{theorem}\cite{VirBloe}
\label{thm3}There exists a unique bounded solution to the problem:
\beq
\left\{\begin{array}{l} \Big[\frac{2}{\beta}\partial_w^2 + (s - w^2)\partial_w + \partial_s\Big]\phi_{\beta}(w,s) = 0 \\ 
\lim_{w \rightarrow -\infty} \phi_{\beta}(w,s) = 0 \\
\lim_{s \rightarrow +\infty} \lim_{w \rightarrow +\infty} \phi_{\beta}(w,s) = 1 \end{array}\right.
\eeq
and $\mathsf{TW}_{\beta}(s) = \lim_{w \rightarrow +\infty} \phi_{\beta}(w,s)$.
\end{theorem}
This can be reformulated in terms of a Schr\"{o}dinger equation with time-dependent potential. Indeed, if we set:
\beq
\psi_{\beta}(w,s) = \phi_{\beta}(w,s)\,\exp\Big[-\frac{\beta}{4}\Big(\frac{w^3}{3} - sw\Big) - h(s)\Big]
\eeq
for a choice of $h(s)$ yet to determine, we find that $\psi_{\beta}(w,s)$ is solution to:
\beq
\label{schro}H_{\beta} \psi_{\beta} = \partial_s \psi_{\beta},\qquad H_{\beta} = -\frac{2}{\beta}\partial_w^2 + U_{\beta}(w,s)
\eeq
with the time-dependent potential:
\beq
\label{pot}U_{\beta}(w,s) = 
\frac{\beta}{8}(w^2-s)^2 +\Big(\frac{\beta}{4} - 1\Big)w - h'(s)
\eeq
However, it is not easy to perform analytic computations with $\mathsf{TW}_{\beta}(s)$ starting from this characterization. One would prefer, for instance, a genuine differential equation characterizing $\mathsf{TW}_{\beta}(s)$ itself.

For $\beta = 2$, it is possible to realize $\psi_{2}(w,s)$ as the solution of a $2 \times 2$ Lax system:
\beq
\label{eqlax} \Psi(w,s) = \left(\begin{array}{c} \psi_{2}(w,s) \\ \overline {\psi}_{2}(w,s) \end{array}\right),\qquad \left\{\begin{array}{l} \partial_w \Psi(w,s) = \mathbf{L}(w,s)\,\Psi(w,s) \\ \partial_s \Psi(w,s) = \mathbf{M}(w,s)\,\Psi(w,s) \end{array}\right.
\eeq
The matrices $\mathbf{L}$ and $\mathbf{M}$ correspond (if $w$ were rescaled by a factor $\frac{1}{2i}$) to the Lax pair of Flaschka and Newell \cite{FN80}:
\beq
\mathbf{L}(w,s) = \left(\begin{array}{cc} \frac{w^2}{2} - \frac{s}{2} - q^2(s) & -wq(s) + q'(s) \\ -wq(s) - q'(s) & -\frac{w^2}{2} + \frac{s}{2} + q^2(s) \end{array}\right), \qquad \mathbf{M}(w,s) = \left(\begin{array}{cc} -\frac{w}{2} & q(s) \\ q(s) & \frac{w}{2} \end{array}\right)
\eeq
if we make the choice $h'(s) = R(s) = \big(q'(s)\big)^2 - sq^2(s) - q^4(s)$. The compatibility condition of the system~\ref{eqlax} is the Painlev\'{e} II equation for $q$, and for any $s$, there exists a unique solution subjected to the asymptotic condition:
\beq
\psi_2(w,s) \in o\Big[\exp\Big\{-\frac{1}{2}\Big(\frac{w^3}{3} - sw\Big)\Big]\Big\}\quad w \rightarrow -\infty
\eeq
and the normalisation $\lim_{s \rightarrow +\infty} \lim_{w \rightarrow +\infty} \psi_2(w,s) = 1$. It can be shown that the Tau function $\tau(s)$ of this system can be represented as $\mathsf{H}(s)^2$ in terms of the Hastings-McLeod solution of Painlev\'{e} II, up to a constant factor (see e.g. \cite{FW01}). This is indeed the representation of $\mathsf{TW}_2(s)$.

However, when $\beta \neq 2$, it is not possible to find a $2 \times 2$ Lax system leading to the Schr\"{o}dinger equation~\ref{schro} for the wave function $\psi_{\beta}(w,s)$. It is also not clear if $\mathsf{TW}_{\beta}(s)$ is a tau function for an appropriate integrable system.


\subsection{Right tail expansion}

In this subsection, we review the known results for the asymptotic behaviour of $\mathsf{TW}_{\beta}(s)$ when $s \rightarrow +\infty$.
The expansion when $s \rightarrow +\infty$ is well known for $\beta=1,2,4$. For a general $\beta >0$, there are much less results. The goal of our paper is to show how to compute recursively all the terms of the large $s$ expansion of $\mathsf{TW}_{\beta}(s)$ for all $\beta >0$.

When $\beta = 1,2$ or $4$, the asymptotic expansion of $\mathsf{TW}_{\beta}(s)$ when $s \rightarrow +\infty$ can be derived in a straightforward way from its representation in terms of solutions of Painlev\'e II (Theorem~\ref{thm1}). We find that:
\bea
\int_s^{\infty} R(t)\,\dd t & = & \frac{e^{-\frac{4s^{3/2}}{3}}}{16\pi\,s^{3/2}}\Big[1 - \frac{35}{2^3\cdot 3}\,s^{-3/2} + \frac{3745}{2^7\cdot 3^2}\,s^{-3} - \frac{805805}{2^{10}\cdot 3^4}\,s^{-9/2} + \frac{289554265}{2^{15}\cdot 3^5}\,s^{-6} \nonumber \\
& & \phantom{\frac{e^{-\frac{4s^{3/2}}{3}}}{16\pi\,s^{3/2}}}\, - \frac{31241084875}{2^{18}\cdot 3^6}\,s^{-15/2} + \frac{23604769513325}{2^{22}\cdot 3^8}\,s^{-9} + O(s^{-21/2})\Big]
\nonumber \\
\int_s^{\infty} q(t)\,\dd t & = & \frac{e^{-\frac{2s^{3/2}}{3}}}{2\sqrt{\pi}\,s^{3/4}}\Big[1 - \frac{41}{2^4\cdot 3}\,s^{-3/2} + \frac{9241}{2^9\cdot 3^2}\,s^{-3} - \frac{5075225}{2^{13}\cdot 3^4}\,s^{-9/2} + \frac{5153008945}{2^{19}\cdot 3^5}\,s^{-6} \nonumber \\
& & \phantom{\frac{e^{-\frac{2s^{3/2}}{3}}}{2\sqrt{\pi},s^{3/4}}}\, - \frac{1674966309205}{2^{23}\cdot 3^6}\,s^{-15/2} + \frac{3985569631633205}{2^{28}\cdot 3^8}\,s^{-9} + O(s^{-21/2})\Big] \nn
\eea

\begin{theorem}
\label{sau}The following asymptotic expansion when $s \rightarrow +\infty$ holds:
\bea
1 - \mathsf{TW}_{1}(s) & = & \frac{e^{-\frac{2s^{3/2}}{3}}}{4\sqrt{\pi}\,s^{3/4}}\Big[1 - \frac{41}{2^4\cdot 3}\,s^{-3/2} + \frac{9241}{2^9\cdot 3^2}\,s^{-3} - \frac{5075225}{2^{13}\cdot 3^4}\,s^{-9/2} + \frac{5153008945}{2^{19}\cdot 3^5}\,s^{-6} \nonumber \\
& & \phantom{\frac{e^{-\frac{2s^{3/2}}{3}}}{2\sqrt{\pi},s^{3/4}}}\, - \frac{1674966309205}{2^{23}\cdot 3^6}\,s^{-15/2} + \frac{3985569631633205}{2^{28}\cdot 3^8}\,s^{-9} + O(s^{-21/2})\Big] \nn \\
1 - \mathsf{TW}_{2}(s) & = & \frac{e^{-\frac{4s^{3/2}}{3}}}{16\pi\,s^{3/2}}\Big[1 - \frac{35}{2^3\cdot 3}\,s^{-3/2} + \frac{3745}{2^7\cdot 3^2}\,s^{-3} - \frac{805805}{2^{10}\cdot 3^4}\,s^{-9/2} + \frac{289554265}{2^{15}\cdot 3^5}\,s^{-6} \nonumber \\
& & \phantom{\frac{e^{-\frac{4s^{3/2}}{3}}}{16\pi\,s^{3/2}}}\, - \frac{31241084875}{2^{18}\cdot 3^6}\,s^{-15/2} + \frac{23604769513325}{2^{22}\cdot 3^8}\,s^{-9} + O(s^{-21/2})\Big] \nonumber
\\
1 - \mathsf{TW}_{4}(s) & = & 
\frac{e^{-\frac{8s^{3/2}}{3}}}{512\pi\,s^{3}}
\Big[1 - \frac{143}{2^4\cdot 3}\,s^{-3/2} + \frac{41509}{2^9\cdot 3^2}\,s^{-3} - \frac{20443229}{2^{13}\cdot 3^4}\,s^{-9/2} + \frac{15418569025}{2^{19}\cdot 3^5}\,s^{-6} \nonumber \\
& & \phantom{\frac{3\,e^{-\frac{4s^{3/2}}{3}}}{32\pi\,s^{3/2}}}\, -\frac{3330409204735}{2^{23}\cdot 3^6}\,s^{-15/2} + \frac{4908974519795465}{2^{28}\cdot 3^8}\,s^{-9} + O(s^{-21/2})\Big] \nn
%
%
\eea
\end{theorem}

When $\beta$ is arbitrary, taking advantage of the connection to a diffusion process
in a time-dependent potential, Dumaz and Vir\'{a}g have established the first two terms of the asymptotic expansion when $s \rightarrow +\infty$:
\begin{theorem}\cite{VirDum}
For $\beta>0$, the following asymptotics hold when $s \rightarrow +\infty$:
\beq
\label{Dumvir}1 - \mathsf{TW}_{\beta}(s) = s^{-\frac{3\beta}{4} + o[(\ln s)^{-1/2}]}\,e^{-\frac{2\beta s^{3/2}}{3}}
\eeq
\end{theorem}
In this article, we study heuristically the asymptotic expansion of $\mathsf{TW}_{\beta}(s)$ when $s \rightarrow +\infty$. Our result takes the form:
\begin{proposition} (Heuristics)
\label{prp}
\beq
\label{qpqp}1 - \mathsf{TW}_{\beta}(s) = \frac{\Gamma\big(\frac{\beta}{2}\big)}{(4\beta)^{\frac{\beta}{2}}\,2\pi}\,s^{-\frac{3\beta}{4}}\,e^{-\frac{2\beta s^{3/2}}{3}}\,\exp\Big[\sum_{m \geq 1} \frac{\beta}{2}\,R_m\big(\frac{2}{\beta}\big)\,s^{-\frac{3m}{2}}\Big]
\eeq
where $R_m$ are polynomials of degree atmost $m + 1$ in $\frac{2}{\beta}$, with rational coefficients, and the first few are:
\bea
R_1(X) & = & \frac{1}{24}(-5X^2 + 9X - 39)  \\
R_2(X) & = & \frac{5}{64}(11 X^2 - 19 X + 36) \\
R_3(X) & = & \frac{1}{4608}(-1105 X^4 + 3240 X^3 - 23325 X^2 + 34938 X - 41433)
\eea
\end{proposition}
\vspace{0.2cm}
In particular, the first two terms are in agreement with the rigorous result of Dumaz and Vir\'{a}g, see Eqn.~\ref{Dumvir}.
For $\beta = 1,2$ and $4$, the orders written above also match with those in Theorem~\ref{sau}. In section \ref{sec:ourstudy}, we present a method to find the $R_m$ recursively. For this purpose, one only needs to write down the solution of a recursive system of linear equations (and no inversion of linear operator is required), and then determine its behavior in some scaling limit. 

\subsection{Left tail expansion}

In this subsection, we recall for completeness the situation 
for the asymptotic behaviour of $\mathsf{TW}_{\beta}(s)$ when $s \rightarrow -\infty$.

When $\beta = 1,2$ or $4$, this asymptotic expansion can be obtained again from the known asymptotics of $q(s)$, except that results about the total integral of $q(s)$ are needed to obtain the constant prefactor. This last step was performed by Baik, Buckingham and diFranco \cite{BBdiF}, and we have:
\begin{theorem}
\bea
\mathsf{TW}_1(s) & = & 2^{-11/48}\,e^{\zeta'(-1)/2}\,\exp\Big[ -\frac{|s|^3}{24} - \frac{|s|^{3/2}}{3\sqrt{2}} - \frac{\ln |s|}{16} - \frac{|s|^{-3/2}}{24\sqrt{2}} \nonumber \\
& & \phantom{2^{-11/48}\,e^{\zeta'(-1)/2}\,\exp}\, + \frac{3|s|^{-3}}{128} - \frac{73|s|^{-9/2}}{1152\sqrt{2}} + \frac{63|s|^{-6}}{512} + O\big(|s|^{-15/2}\big)\Big] \nn \\
\mathsf{TW}_2(s) & = & 2^{1/24}\,e^{\zeta'(-1)}\,\exp\Big[-\frac{|s|^3}{12} - \frac{\ln |s|}{8} + \frac{3|s|^{-3}}{64} + \frac{63|s|^{-6}}{256} + O\big(|s|^{-9}\big)\Big] \nn\\
\mathsf{TW}_4(s) & = & 2^{-37/48}\,e^{\zeta'(-1)/2}\,\exp\Big[ -\frac{|s|^3}{6} + \frac{|s|^{3/2} \sqrt{2}}{3} - \frac{\ln |s|}{16} + \frac{|s|^{-3/2}}{48\sqrt{2}} \nonumber \\
& & \phantom{2^{-37/48}\,e^{\zeta'(-1)/2}}\, {+ \frac{3|s|^{-3}}{512} +
\frac{73|s|^{-9/2}}{9216 \sqrt{2}} + \frac{63|s|^{-6}}{8192} + O\big(|s|^{-15/2}\big)} \Big] \nonumber
\eea
\end{theorem}
Let us mention that for $\beta = 2$, the value $2^{1/24}e^{\zeta'(-1)}$ of the constant prefactor has been conjectured originally by Tracy and Widom \cite{TW92} and first proved by Deift, Its and Krasovsky \cite{DIK}. Notice that only integer powers of $|s|$ appear in the expansion of $\ln \mathsf{H}(s)$ when $s \rightarrow -\infty$. Moreover, $\mathsf{E}(s)$ is of order $\exp(-\frac{|s|^{3/2}}{3\sqrt{2}})$ when $s \rightarrow -\infty$ according to \cite[Corollary 1.3]{BBdiF}, so is exponentially small compared to the term $\mathsf{E}^{-1}(s)$ in $\mathsf{TW}_{4}(s)$. For this reason, the asymptotics to all orders of $\mathsf{TW}_4(s)$ are the same as that of $\mathsf{TW}_1(s)$ except\footnote{The coefficients in the asymptotic expansion of $\textsf{F}_{4}(s) = \mathsf{TW}_{4}(2^{2/3}s)$, which is the usual definition of the Tracy-Widom function for $\beta = 4$, also differ by powers of $2$.} for a sign change in front of each half-integer power of $|s|$.

In a previous work \cite{BEMN}, we have shown heuristically that, for all $\beta>0$,
 $\mathsf{TW}_{\beta}(s)$ admits a full asymptotic expansion when $s \rightarrow -\infty$, of the form:
\beq
 \mathsf{TW}_{\beta}(s) = \tau_{\beta}\,\exp\Big[ -\frac{\beta |s|^3}{24} + \frac{(\beta - 2)|s|^{3/2}}{3\sqrt{2}} + \frac{\frac{\beta}{2} + \frac{2}{\beta} - 3}{8}\,\ln |s| + \sum_{m \geq 1} \frac{\frac{\beta}{2}\,L_m\big(\frac{2}{\beta}\big)\,|s|^{-\frac{3m}{2}}}{2^{\frac{7m}{2} + 2}\cdot 3\cdot m}\Big] \nonumber
 \eeq
where $L_{m}$ are polynomials of degree $m + 2$ in $\frac{2}{\beta}$, with integer coefficients. $L_m$ is reciprocal up to a sign:
\beq
L_m\big(\frac{\beta}{2}\big) = (-1)^m\,\big(\frac{\beta}{2}\big)^m\,L_m\big(\frac{2}{\beta}\big)
\eeq
The $L_m$ can be computed recursively (although we do not have a straightforward relation of the form $L_m = f_m(L_{m - 1},\ldots,L_1)$), their coefficients are related to the symplectic invariants introduced by Chekhov and Eynard \cite{CE06}, computed for the curve of equation $y^2 = x + \frac{1}{x} - 2$. The first few are:
\bea
L_1(X) & = &  -54 + 193 X - 193 X^2 + 54 X^3 \nonumber \\
L_2(X) & = & 2(-1190 + 5247 X - 8042 X^2 + 5247 X^3 - 1190 X^4) \nn \\
L_3(X) & = & -119862 + 608591 X - 1199970 X^2 + 1199970 X^3 - 608591 X^4 + 119862 X^5 \nn \\
L_4(X) & = & 2(- 3467274 + 19186065 X - 44032530 X^2 + 56724246 X^3 - 44032530 X^4 \nonumber \\
& & \phantom{2\,} + 19186065 X^5 - 3467274 X^6) \nonumber
\eea 
The constant prefactor is given by:
\beq
\tau_{\beta} = 2^{\frac{-25 \frac{2}{\beta} + 51 - 25 \frac{\beta}{2}}{24}}\,\sqrt{2\pi}\,e^{\chi'(0 ; \frac{2}{\beta})}
\eeq
where $\chi(s;\alpha)$ is a double zeta function, namely the analytic continuation of the series defined for $\mathrm{Re}\,s > 2$:
\beq
\chi(s ; \alpha) = \sum_{\substack{m,m ' \geq 0 \\ (m,m') \neq (0,0)}} \frac{1}{(m + \alpha m')^{s}},\qquad \chi(s ; \alpha^{-1}) = \alpha^{s}\,\chi(s ; \alpha)
\eeq
Moreover, the reciprocity of $L_m$ implies a duality valid to all orders in the asymptotic expansion when $s \rightarrow -\infty$:
\beq
\mathsf{TW}_{\beta}(s) = \big(\frac{\beta}{2}\big)^{\frac{1}{6}(\frac{2}{\beta} + \frac{\beta}{2}) - 1}\,\widetilde{\mathsf{TW}}_{\frac{4}{\beta}}\big[\big(\frac{2}{\beta}\big)^{2/3}\! s\big]
\eeq
where the tilde indicates that we take the other sign for squareroots of $|s|$ in the expansion. Those heuristic results are in agreement with all previously known results on the Tracy-Widom laws. In \cite{BEMN}, we have also argued that the coefficients of this asymptotic expansion are universal, up to a non universal normalization of $s$. 
The universality class is defined by the blow-up of the spectral density in presence of hard wall close to the edge of the eigenvalue spectrum, which is generically the curve of equation $\widehat{y}^2 = \widehat{x} + \widehat{x}^{-1} + 2$.

The present article completes this work by studying with similar methods the right tail asymptotics of $\mathsf{TW}_{\beta}(s)$ (i.e. the limit $s\rightarrow +\infty$). We postpone an argument for universality of the right tail expansion to a future version of this article.
 
\subsection{Method}
\label{subsec:TWfromLD}

In this subsection, we explain how we can recover the right asymptotics (ie limit $s\to\infty$)
of the Tracy-Widom laws $\mathsf{TW}_{\beta}(s)$ from the right large deviations of the maximal
eigenvalue of the $\mathrm{G}\beta\mathrm{E}$.

We first study the large deviations of $\lambda_{\mathrm{max}}$, i.e. the behavior when $N \rightarrow +\infty$ of the probability density function of $\lambda_{\rm max}$:
\beq
\mathcal{G}_{N,\beta}^{V}(a) =  -\partial_a \mu_{N,\beta}^{V}[\lambda_{\mathrm{max}} > a]
\eeq
when $a$ is independent of $N$ and is located to the right of the bulk spectrum. Deviations of $\lambda_{\mathrm{max}}$ 
of order one from 
its mean value $a^* = \lim_{N \rightarrow \infty} \mu_{N,\beta}^{V}[\lambda_{\mathrm{max}}]$ are rare events when $N$ is large: the probability $\mathcal{G}_{N,\beta}(a)$ is actually exponentially small. Under some assumptions, we will show that $\mathcal{G}_{N,\beta}(a)$ admits a large $N$ asymptotic expansion of the form:
\beq
\label{exp}\mathcal{G}_{N,\beta}^{V}(a) = C_{N,\beta}\,\exp\Big(\sum_{k \geq 0} N^{1 - k}\,\mathcal{G}^{[k]}_{\beta}(a)\Big)
\eeq 
Then, if we substitute $a = a^* + N^{-2/3}s$ in $\mathcal{G}^{[k]}_{\beta}(a)$,
ie if we zoom close to the mean value (approaching it from the right)
looking at fluctuations of order $N^{-2/3}$ from the mean value $a^*$
(instead of order one), we will find that:
\beq
\mathcal{G}^{[k]}_{\beta}(a = a^* + N^{-2/3}s) \sim N^{k - 1}\,\widehat{\mathcal{G}}^{[k]}_{\beta}(s)
\eeq
This means that all terms in Eqn.~\ref{exp} become of order $O(1)$, which is not surprising since $\mathcal{G}_{N,\beta}(a^* + sN^{-2/3})$ must be of order $1$ and not exponentially small when $N$ is large. 
We indeed expect that $\mu_{N,\beta}^{V}[\lambda_{\mathrm{max}} \leq a]=1-\mu_{N,\beta}^{V}[\lambda_{\mathrm{max}} > a]$ for $a=a^* + sN^{-2/3}$ converges to the Tracy-Widom law $\mathsf{TW}_{\beta}(s)$. Actually, we will see that a stronger result holds:
\beq
\label{gk}\forall k \geq 3,\,\qquad \widehat{\mathcal{G}}^{[k]}_{\beta}(s) = M_k\left(\frac{\beta}{2}\right)\,(s^{3/2})^{1-k}
\eeq
where $M$ is a Laurent polynomial in $\frac{\beta}{2}$, with rational coefficients.

This suggests, and this is the non rigorous point in our derivation, to interprete $\exp\Big(\sum_{k \geq 0} M_k\big(\frac{\beta}{2}\big)\,s^{\frac{3}{2}(1 - k)}\Big)$ as the asymptotic expansion of:
\beq
 \mathsf{TW}'_{\beta}(s) = \lim_{N \rightarrow +\infty} N^{-2/3} \,\mathcal{G}_{N,\beta}(a = a^* + N^{-2/3}s)
\eeq
Notice that Eqn.~\ref{gk} holds only for $k \geq 3$. The terms for $k = 0,1,2$ have to be handled separately, and the details of our computations show that they combine with the large $N$ behavior of the constant $N^{-2/3}\,C_{N,\beta}$ to give a finite result, which reproduces Eqn.~\ref{Dumvir} as well as a constant prefactor $\widehat{\tau}_{\beta}$ which agrees with the values known for $\beta = 1,2$ and $4$.

\section{Large deviations}
\label{sec:ourstudy}

In this section, we study the right large deviations of the maximal eigenvalue $\lambda_{\rm max}$ in the $G\beta E$, i.e. the maximum of the $\lambda_i$ where the $\lambda_i$
are random variables distributed according to the law in Eqn.~\ref{mes}
with a quadratic potential $V(x)=\frac{x^2}{2 t}$. We keep $t>0$ arbitrary in the calculations, so that the result can be read directly with the various conventions of normalization found in the literature. Our approach is based on the so-called Schwinger-Dyson equations for correlators of eigenvalues. 

\subsection{Beta ensembles and Schwinger-Dyson equations}

We want to compute $\mathcal{G}_{N,\beta}^{V}(a)$, the probability density function
 of $\lambda_{\mathrm{max}}$, with respect to the measure defined in Eqn.~\ref{mes}:
\bea
\mathcal{G}_{N,\beta}^{V}(a) & = & -\partial_a\mu_{N,\beta}^{V}[\lambda_{\mathrm{max}} > a] \nonumber \\
& = & \frac{N\,e^{-\frac{N\beta}{2}\,V(a)}}{Z_{N,\beta}^{V}}\,\int_{]-\infty,a]^{N - 1}} \prod_{i = 1}^{N - 1}\dd\lambda_i\,e^{-\frac{N\beta}{2}\,V(\lambda_i)}\,(a - \lambda_i)^{\beta}\,\prod_{1 \leq i < j \leq N - 1} |\lambda_i - \lambda_j|^{\beta} \nonumber \\
& = & \frac{N\,e^{-\frac{N\beta}{2}\,V(a)}\,Z_{N - 1,\beta}^{V_{N,a}}}{Z_{N,\beta}^{V}} \label{eq:GNZN}
\eea
where:
\beq
V_{N,a}(x) = \left\{\begin{array}{lll} \frac{NV(x)}{N - 1} - \frac{2 \ln(a - x)}{N - 1} & & \mathrm{if}\,\,x < a \\ +\infty & & \mathrm{if}\,\,x > a \end{array}\right.
\eeq
and where for an integer $M$ and a function $U$ the partition function $Z_{M,\beta}^{U}$
is defined as:
\beq
Z_{M,\beta}^{U}=  \prod_{i = 1}^M\int_{-\infty}^{\infty} \dd\lambda_i\,e^{-\frac{M\beta}{2}\,U(\lambda_i)}\,\prod_{1 \leq i < j \leq M} |\lambda_i - \lambda_j|^{\beta}
\eeq

In other words, $\mathcal{G}_{N,\beta}^{V}(a)$ can be computed in terms of the partition function $Z_{N - 1,\beta}^{V_{N,a}}$
of a $\beta$ ensemble with $N - 1$ eigenvalues in a modified potential $V_{N,a}$ depending on $N$ and $a$.
\\

For any probability measure defined as:
\beq
\dd \mu_{M,\beta}^{U}(\lambda_1,\ldots,\lambda_M) = \frac{1}{Z_{M,\beta}^{U}}\,\prod_{i = 1}^M \dd\lambda_i\,e^{-\frac{M\beta}{2}\,U(\lambda_i)}\,\prod_{1 \leq i < j \leq M} |\lambda_i - \lambda_j|^{\beta}
\eeq
 we can derive Schwinger-Dyson equations, which give relations between some expectation values with respect to $\mu_{M,\beta}^{U}$, by using integration by parts. To write them down, it is convenient to introduce the correlators:
\beq
W_n(x_1,\ldots,x_n) = \mu_{M,\beta}^{U}\Big[\Big(\prod_{j = 1}^n \sum_{i_j = 1}^{M} \frac{1}{x_j - \lambda_{i_j}}\Big)_c\Big]
\eeq
where $c$ stands for cumulant and $\mu_{M,\beta}^{U}[\cdots]$ denotes the mean value
of $\cdots$ with respect to $\mu_{M,\beta}^{U}$. The $W_n$ are related to $n$-point correlations of density of the $\lambda_i$'s. For instance, $W_1$ is related to the spectral density
\beq\label{specdens}
\rho(x)\equiv\mu_{M,\beta}^{U}\Big[\frac{1}{M}\sum_{i = 1}^M \delta(x - \lambda_i)\Big] = \lim_{\epsilon \rightarrow 0^+} \frac{W_1(x - i\epsilon) - W_1(x + i\epsilon)}{2i\pi\,M}
\eeq
for $x \in \mathbb{R}$, in the sense of distributions. Let us also define:
\beq
P_n(x;x_2,\ldots,x_n) = \mu_{M,\beta}^{U}\Big[\Big(\Big(\sum_{i = 1}^{M} \frac{U'(x) - U'(\lambda_i)}{x - \lambda_i}\Big)\prod_{j = 2}^n \sum_{i_j = 1}^M \frac{1}{x_j - \lambda_{i_j}}\Big)\Big)_c\Big]
\eeq
In those notations, it is implicit that $W_n$ and $P_n$ depend on $M$, $\beta$ and $U$. We also use the notation $x_I = (x_i)_{i \in I}$ when $I$ is a set. The Schwinger-Dyson equations
(also called loop equations) for the beta ensemble have been established many times \cite{CE06,BG11}. We give them here without proof:
\begin{theorem}
For any $x \in \mathbb{R}\setminus U^{-1}(\{+\infty\})$:

\begin{small}
\beq
\label{28}W_2(x,x) + \big(W_1(x)\big)^2 + \Big(1 - \frac{2}{\beta}\Big)(\partial_x W_1)(x) - M\,U'(x)\,W_1(x) + M\,P_1(x) = 0
\eeq
\end{small}

\noindent $\!\!$And for $n \geq 2$, if $I$ denotes the set $\{2,\ldots,n\}$, and $x,x_2,\ldots,x_n \in \mathbb{R}\setminus U^{-1}(\{+\infty\})$ are $n$ points distinct two by two:

\begin{small}\bea
W_{n + 1}(x,x,x_I) + \sum_{J \subseteq I} W_{|J| + 1}(x,x_J)\,W_{n - |J|}(x,x_{I\setminus J}) + \Big(1 - \frac{2}{\beta}\Big)(\partial_x W_n)(x,x_I) & &  \\
- M\,U'(x)\,W_n(x,x_I) + M\,P_n(x,x_I) + \frac{2}{\beta} \sum_{i \in I} \partial_{x_i}\Big(\frac{W_{n - 1}(x,x_{I\setminus\{i\}}) - W_{n - 1}(x_I)}{x - x_i}\Big) & = & 0 \nonumber
\eea
\end{small}

\end{theorem}
We stress that those relations are exact for any finite $M$. Unless explicitly stated, we shall apply these relations to the measure $\mu_{N - 1,\beta}^{V_{N,a}}$ only, i.e. we only use correlators with respect to the modified measure $\mu_{N - 1,\beta}^{V_{N,a}}$.
\\

If we know the correlators, we may come back to the derivative of the partition function.  Since the integration weight in $\mu_{N - 1,\beta}^{V_{N,a}}$ vanishes when one of the $\lambda_i$ equals $a$, we have:
\beq
\partial_a \ln Z_{N - 1,\beta}^{V_{N,a}} = \beta\,W_1(a)
\eeq
We then need to integrate the above expression to get $ \ln Z_{N - 1,\beta}^{V_{N,a}}$
and then the probability density function of $\lambda_{\rm max}$, ie $\mathcal{G}_{N,\beta}^{V}(a)$
(cf Eqn. \eqref{eq:GNZN}).
To obtain the integration constant, let us remark that for any fixed $N$:
\beq
Z_{N - 1}^{V_{N,a}} \mathop{\sim}_{a \rightarrow +\infty} a^{(N - 1)\beta}\,Z_{N - 1,\beta}^{\frac{NV}{N - 1}}
\eeq
Accordingly we get an explicit expression for the probability density function of 
$\lambda_{\rm max}$, ie $\mathcal{G}_{N,\beta}^{V}(a)$ as a function of the first correlator $W_1$:
\beq
\label{cte} \mathcal{G}_{N,\beta}^{V}(a) = \frac{N\,e^{-\frac{N\beta}{2}\,V(a)}\,Z_{N - 1,\beta}^{\frac{NV}{N - 1}}}{Z_{N,\beta}^{V}}\,a^{(N - 1)\beta}\,\exp\Big[-\beta \int_{a}^{\infty}\dd a'\Big(W_1(a') - \frac{N - 1}{a'}\Big)\Big]
\eeq
So, if we want to compute $\mathcal{G}_{N,\beta}^{V}(a)$, our first goal is to compute $W_1(a)$, and for this task we can use Schwinger-Dyson equations.

\subsection{$1/N$ expansion and large deviations}

Under reasonable assumptions on $V$ (which carry on to $V_{N,a}$), one of the author and Guionnet showed in \cite{BG11} that the correlators with respect to $\mu_{N - 1,\beta}^{V_{N,a}}$ admit an asymptotic expansion when $N\rightarrow +\infty$, of the form:
\beq
\label{exo} W_n(x_1,\ldots,x_n) = \sum_{k \geq 0} N^{2 - n - k}\,W_n^{[k]}(x_1,\ldots,x_n)
\eeq
when $x_1,\ldots,x_n$ are points (not necessarily distinct) in a neighborhood of $\infty$ which does not depend on $N$. The important feature in Eqn.~\ref{exo} is that $W_n$ is a $O(N^{2 - n})$. Let us plug this expansion in the Schwinger Dyson equation. But notice first that:
\beq
(N - 1)\,V_{N,a}(x) = N\,V(x) - 2\,\ln(a - x)
\eeq 
allows us to decompose:
\beq
\label{expan} (N - 1)\,P_n(x;x_2,\ldots,x_n) = N\,Q_n(x;x_2,\ldots,x_n) - \frac{2\,W_1(a)}{x - a}
\eeq
where:
\beq
Q_n(x;x_2,\ldots,x_n) = \mu_{N - 1,\beta}^{V_{N,a}}\Big[\Big(\Big(\sum_{i = 1}^{N - 1} \frac{V'(x) - V'(\lambda_i)}{x - \lambda_i}\Big)\prod_{j = 2}^n \Big(\sum_{i_j = 1}^{N - 1} \frac{1}{x_j - \lambda_{i_j}}\Big)\Big)_c\Big]
\eeq
The large $N$ expansion for the correlators also implies an asymptotic expansion for $Q_n$, of the form:
\beq
Q_n(x;x_2,\ldots,x_n) = \sum_{k \geq 0} N^{2 - n - k}\,Q_n^{[k]}(x;x_2,\ldots,x_n)
\eeq

At leading order for $W_1$, we find:
\beq\label{eq:w10}
\big(W_1^{[0]}(x)\big)^2 - V'(x)\,W_1^{[0]} + Q^{[0]}_1(x) = 0
\eeq
which implies that $W_1^{[0]}$ does not depend on $\beta$ and $a$: it coincides with the leading order of the correlator with respect to $\mu_{N,\beta = 2}^{V}$, i.e. of the measure for a random hermitian matrix with potential $V$. For the subleading orders, we find:
\bea
& & \big(V'(x) - 2W_1^{[0]}(x)\big)W_1^{[k]}(x) - Q_1^{[k]}(x) \nonumber \\
& = & W_2^{[k - 2]}(x,x) + \sum_{k' = 1}^{k - 1} W_1^{[k']}(x)\,W_1^{[k - k']}(x) \nonumber \\
\label{eq1}& & + \Big(1 - \frac{2}{\beta}\Big)(\partial_x W_1^{[k - 1]})(x) + 2\,\frac{W_1^{[k - 1]}(x) - W_1^{[k - 1]}(a)}{x - a} 
\eea
and for $n \geq 2$:
\bea
& & \big(V'(x) - 2W_1^{[0]}(x)\big)W_n^{[k]}(x,x_I) - Q_n^{[k]}(x;x_I) \nonumber \\
& = & W_{n + 1}^{[k - 2]}(x,x,x_I) + \sum_{\substack{J \subseteq I\quad 0 \leq k' \leq k \\ (J,k') \neq (\emptyset,0),(I,k')}} W_{|J| + 1}^{[k]}(x,x_J)\,W_{n - |J|}^{[k - k']}(x,x_{I\setminus J}) \nonumber \\
& & + \Big(1 - \frac{2}{\beta}\Big)(\partial_x W_n^{[k - 1]})(x,x_I) + 2\,\frac{W_n^{[k - 1]}(x) - W_n^{[k - 1]}(a)}{x - a} \nonumber \\
\label{eq2} & & + \frac{2}{\beta}\sum_{i \in I} \partial_{x_i}\Big(\frac{W_{n - 1}^{[k]}(x,x_{I\setminus\{i\}}) - W_{n - 1}^{[k]}(x_I)}{x - x_i}\Big)
\eea
In fact, $Q_n(x;x_I) = (\mathcal{O}\cdot W_n)(x,x_I)$ is found by action of a linear operator $\mathcal{O}$ on $W_n$. Provided one knows how to invert $(V' - 2W_1^{[0]}\big)\mathrm{Id} + \mathcal{O}$, the relations above form a triangular system which can be solved recursively to find $W_n(x_1,\ldots,x_n)$. The recursion is on the integer $n + k$.

\subsection{The Gaussian case}
\label{subsec:corrgauss}

When $V$ is a quadratic potential, i.e. $V(x) = \frac{x^2}{2t}$, it is even simpler, because all the quantities $Q_n$ are known a priori:
\beq
\label{Qeq}Q_1(x) = \delta_{n,1}\,\frac{N - 1}{t},\qquad Q_n(x;x_2,\ldots,x_n) = 0\,\,\mathrm{for}\,\,n \geq 2
\eeq

Moreover, when $a > 2\sqrt{t}$ ie $a>a^*$
(for $V(x) = \frac{x^2}{2t}$ the mean value of $\lambda_{\rm max}$
is $a^*=2 \sqrt{t}$ when $N\to\infty$),
 the effective potential $V_{N,a}^{\mathrm{G}\beta\mathrm{E}}$ satisfies the assumptions:
\begin{itemize}
\item[$\diamond$] $V_{N,a}^{\mathrm{G}\beta\mathrm{E}}\,:\,]-\infty,a] \rightarrow \mathbb{R}$ is continuous, and admits a $1/(N - 1)$ expansion.
\item[$\diamond$] its large $(N - 1)$ limit is $V^{
\mathrm{G}\beta\mathrm{E}}(x) = \frac{x^2}{2t}$, and it is known that the large $(N - 1)$ spectral density with respect to $\mu_{N - 1,\beta}^{\mathrm{G}\beta\mathrm{E}}$ has a connected support $I_t = [-2\sqrt{t},2\sqrt{t}]$, and behaves as a squareroot $\pm 2\sqrt{t}$.
\item[$\diamond$] When $a > 2\sqrt{t}$, $V_{N,a}$ is analytical in a neighborhood of $I_t$.
\end{itemize}
According to \cite[Proposition 1.1]{BG11}, this ensures the existence of a $1/(N - 1)$ expansion for the partition function and the correlators, which can be repackaged into a $1/N$ expansion as stated above in Eqn.~\ref{exo}.

Eqn.~\ref{Qeq} imply that, for the Gaussian potential, Eqns.~\ref{eq1} and \ref{eq2} really form a recursion on $n + k$ without any linear operator to be inverted. The leading order of $W_1$ (cf Eqn.~\ref{eq:w10}):
\beq \label{eq:W10gauss}
W_1^{[0]}(x) = \frac{x - \sqrt{x^2 - 4t}}{2t}
\eeq
is associated to a large $N$ spectral density $\rho(x)$ given by
(cf Eqn. \eqref{specdens}):
\beq
\rho(x) = \frac{\sqrt{x^2 - 4t}}{2\pi t}
\eeq 
We recover without surprise the Wigner semi-circle law, which is valid for all values of $\beta > 0$ \cite{Wig58,Johan}. Let us write $Y(x) = V'(x) - 2W_1^{[0]}(x) = \frac{1}{t}\sqrt{x^2 - 4t}$. To compute the subleading orders, it is convenient to introduce a uniformization variable: $z$, which maps $x \in \mathbb{C}\setminus [-2\sqrt{t},2\sqrt{t}]$ to $\mathbb{C}\setminus \overline{D}(0,1)$:
\beq
x = \sqrt{t}\Big(z + \frac{1}{z}\Big)\quad \Leftrightarrow \quad z = \frac{x + \sqrt{x^2 - 4t}}{2\sqrt{t}}
\eeq
and to work with modified correlators:
\beq
\omega_n^{[k]}(z_1,\ldots,z_n) = x'(z_1)\cdots x'(z_n)\,W_n(x(z_1),\ldots,x(z_n))
\eeq
These quantities are now rational functions of $z_1,\ldots,z_n$. Let us call $\alpha$ such that $x(\alpha) = a$ and $|\alpha| > 1$. One can prove by recursion that the $\omega_n^{[k]}$
have poles only at $z_i = 1,-1,\alpha^{-1},0$ and $z_i = z_j^{-1}$ for $i \neq j$.

From Eqn.~\ref{eq:W10gauss}, we have $\omega_1^{[0]}(z) = (z^{-1} - z^{-3})$, and we find at the first step of the recursion, cf. Eqn.~\ref{eq1}:
\bea
\label{eq11} \omega_1^{[1]}(z) & = & \frac{\frac{2}{\beta} - 1}{2}\Big(\frac{1}{z - 1} + \frac{1}{z + 1}\Big) + \frac{-\frac{2}{\beta} + 2}{z} - \frac{2}{z - \alpha^{-1}} \\
\label{eq20} \omega_2^{[0]}(z_1,z_2) & = &  \frac{2}{\beta}\,\frac{1}{(z_1z_2 - 1)^2} 
\eea
Then, at the second step, cf. Eqn.~\ref{eq2}:
\begin{footnotesize}
\bea
\omega_1^{[2]}(z) & = & \frac{2\alpha\Big(- \frac{2}{\beta} + 1\Big)}{(\alpha^2 - 1)^2(z - \alpha^{-1})^{2}} \\
& & + \frac{5\big(\frac{2}{\beta}\big)^2 - 9\,\frac{2}{\beta} + 5}{16(z - 1)^4} + \frac{\big(\frac{2}{\beta}\big)^2 + \frac{9\alpha + 7}{\alpha - 1}\big(- \frac{2}{\beta} + 1\big)}{16(z - 1)^3} + \frac{-\big(\frac{2}{\beta}\big)^2 + \frac{2}{\beta} + \frac{7\alpha^2 + 18\alpha + 7}{(\alpha - 1)^2}}{32(z - 1)^2} \nonumber \\
 & & \frac{-5\big(\frac{2}{\beta}\big)^2 + 9\,\frac{2}{\beta} - 5}{16(z + 1)^4} + \frac{\big(\frac{2}{\beta}\big)^2 + \frac{9\alpha - 7}{\alpha + 1}\big(- \frac{2}{\beta} + 1\big)}{16(z + 1)^3} + \frac{\big(\frac{2}{\beta}\big)^2 - \frac{2}{\beta} + \frac{-7\alpha^2 + 18\alpha - 7}{(\alpha + 1)^2}}{32(z + 1)^2} \nonumber \\
\omega_2^{[1]}(z_1,z_2) & = & 
\frac{2}{\beta}\Big(\frac{2}{\beta} - 1\Big)\Big[\frac{\frac{1}{2}}{(z _1- 1)^3(z_2 - 1)^2} - \frac{\frac{1}{2}}{(z _1+ 1)^3(z_2 + 1)^2} 
 \\
& & \phantom{
\frac{2}{\beta}\Big(\frac{2}{\beta} - 1\Big)} + \frac{2\,z_2^2}{(z_1z_2 - 1)^3(z_2^2 - 1)^2} + \frac{z_2^2(1 + 3z_2^2)}{(z_1z_2 - 1)^2(z_2^2 - 1)^3}\Big] \nonumber \\
& & + \frac{\frac{1}{2}\,\frac{2}{\beta}\frac{1 + \alpha}{1 - \alpha}}{(z_1 - 1)^2(z_2 - 1)^2} + \frac{\frac{1}{2}\,\frac{2}{\beta}\frac{1 - \alpha}{1 + \alpha}}{(z_1 + 1)^2(z_2 + 1)^2} \nonumber \\
\omega_3^{[0]}(z_1,z_2,z_3) & = & \frac{8}{\beta^2}\,\frac{(1 + z_1z_2 + z_2z_3 + z_3z_1)(z_1 + z_2 + z_3 + z_1z_2z_3)}{
(z_1^2 - 1)^{2}(z_2^2 - 1)^{2} (z_3^2 - 1)^{2}}
\eea
\end{footnotesize}
$\!\!$and so on. Despite its non-symmetric expression, $\omega_2^{[1]}(z_1,z_2)$ is actually symmetric in $z_1,z_2$, and in general $\omega_n^{[k]}(z_1,\ldots,z_n)$ must be symmetric functions of all their variables. 

Actually, we remark that the auxiliary quantity $P_n(x;x_I)$, appearing in the Schwinger-Dyson equations \ref{eq1} and \ref{eq2}, is a priori explicitly known in all the classical ensemble, i.e. Gaussian, Laguerre, and Jacobi ensembles. This simplification makes the approach based on Schwinger-Dyson equations attractive to study large $N$ expansions in those ensembles.

\subsection{Large deviations of the maximal eigenvalue in the Gaussian case}
\label{sub1}

Using Eqn.~\ref{cte} and the expressions obtained recursively above for $\omega_1^{[m]}(z)$, we can then find
after integration the desired asymptotic expansion (Eqn.~\ref{exp}) of
the probability density function of $\lambda_{\rm max}$, ie $\mathcal{G}_{N,\beta}^{\mathrm{G}\beta\mathrm{E}}(a)$. 
The prefactor in Eqn.~\ref{cte} can be computed exactly using Selberg's integrals:

\begin{footnotesize}
\bea
\mathcal{G}_{N,\beta}^{\mathrm{G}\beta E}(a) & = & \frac{\Gamma\big(\frac{\beta}{2}\big)}{\sqrt{2\pi}\,\Gamma\big(1 + \frac{N\beta}{2}\big)}\Big(\frac{N\beta}{2}\Big)^{\frac{\beta N + 3 - \beta}{2}}\,t^{\frac{\beta - 1 - \beta N}{2}}\,a^{(N - 1)\beta} \\
& & \times \exp\Big(-\frac{N\beta}{2}\,\frac{a^2}{2t}\Big)\,\exp\Big[ - \beta \sum_{k \geq 0} N^{1 - k} \int_{\alpha(a)}^{\infty} \dd\alpha'\Big(\omega_1^{[k]}(\alpha') - \frac{(\delta_{k,0} - \delta_{k,1})(\alpha'^2 - 1)}{\alpha'(\alpha'^2 + 1)}\Big)\Big] \nonumber
\eea
\end{footnotesize}

\noindent $\!\!$where we recall $\alpha(a) = \frac{a + \sqrt{a^2 - 4t}}{2\sqrt{t}}$, ie
$a=\sqrt{t}\left(\alpha+\frac{1}{\alpha}\right)$. More explicitly, using Stirling's formula:
\begin{proposition}
\label{expo} $\mathcal{G}_{N,\beta}^{\mathrm{G}\beta\mathrm{E}}(a) = - \partial_a\mu_{N,\beta}^{\mathrm{G}\beta\mathrm{E}}[\lambda_{\mathrm{max}} > a]$, for $a > 2\sqrt{t}$, has a large $N$ expansion of the form:
\begin{footnotesize}
\bea
\mathcal{G}_{N,\beta}^{\mathrm{G}\beta\mathrm{E}}(a) & = & \frac{N^{1 - \frac{\beta}{2}}}{\sqrt{t}}\,\frac{\big(\frac{\beta}{2}\big)^{1 - \frac{\beta}{2}}\,\Gamma\big(\frac{\beta}{2}\big)}{2\pi}\,\exp\Big\{
-\beta N\left(
\frac{a}{4 t}\sqrt{a^2-4 t}+\ln\left(
\frac{a-\sqrt{a^2-4 t}}{2\sqrt{t}}
\right)
\right)
%
%
\nonumber \\
& & +\frac{1}{2}\left(1-\frac{3\beta}{2}\right)\ln \left(\frac{a^2-4 t}{t}\right) + \left(\frac{\beta}{2}-1\right)
\ln\left(\frac{a+\sqrt{a^2-4 t}}{2\sqrt{t}}\right) \nonumber \\
& & + \sum_{m \geq 1} N^{-m}\,\Big(-\frac{B_{m + 1}}{m(m + 1)}\,\left(\frac{2}{\beta}\right)^{m} -  \beta\int_{\alpha(a)}^{\infty} \dd\alpha'\,\omega_1^{[m + 1]}(\alpha')\Big)\Big\}
\eea
\end{footnotesize}

\noindent $\!\!$where $B_{m + 1}$ are the Bernoulli numbers (and they vanish when $m + 1$ is odd). 
\end{proposition}

\noindent $\!\!$This large $N$ asymptotic expansion describes the probability of a large deviation of $\lambda_{\mathrm{max}}$ to the right of the edge of the spectrum. The leading order is in agreement with the result of Majumdar and Vergassola who computed it using a Coulomb gas method \cite{MajVer}, and the first correction has been found recently for $\beta = 2$ by Majumdar and one of the authors using a method of orthogonal polynomials on a semi-infinite interval \cite{NadMaj}. We give below the three first decaying corrections:

\begin{footnotesize}
\bea
 -\int_{\alpha(a)}^{\infty} \dd\alpha'\,\omega_1^{[2]}(\alpha') 
& = & \frac{-5\big(\frac{2}{\beta}\big)^2 + 27\,\frac{2}{\beta} - 39}{6(\alpha^2 - 1)^3} + \frac{-3\big(\frac{2}{\beta}\big)^2 + 19\,\frac{2}{\beta} - 33}{4(\alpha^2 - 1)^2} + \frac{\frac{2}{\beta} - 4}{2(\alpha^2 - 1)}  \\
-\int_{\alpha(a)}^{\infty} \dd\alpha'\,\omega_1^{[3]}(\alpha') & = &  \frac{-10\big(\frac{2}{\beta}\big)^3 + 73\big(\frac{2}{\beta}\big)^2 - 191\,\frac{2}{\beta} + 180}{2(\alpha^2 - 1)^6} \nonumber \\
& & + \frac{-25\big(\frac{2}{\beta}\big)^3 + 187\big(\frac{2}{\beta}\big)^2 - 507\,\frac{2}{\beta} + 501}{2(\alpha^2 - 1)^5} + \frac{-80\big(\frac{2}{\beta}\big)^3 + 627\big(\frac{2}{\beta}\big)^2 - 1807\,\frac{2}{\beta} + 1926}{8(\alpha^2 - 1)^4}  \nonumber \\
& & + \frac{-15\,\big(\frac{2}{\beta}\big)^3 + 133\big(\frac{2}{\beta}\big)^2 - 438\,\frac{2}{\beta} + 539}{6(\alpha^2 - 1)^3} + \frac{3\big(\frac{2}{\beta}\big)^2 - 20\,\frac{2}{\beta} + 38}{4(\alpha^2 - 1)^2} \\
 -\int_{\alpha(a)}^{\infty} \dd\alpha'\,\omega_1^{[4]}(\alpha') & = & \frac{-1105\big(\frac{2}{\beta}\big)^4 + 9720\big(\frac{2}{\beta}\big)^3 - 34557\big(\frac{2}{\beta}\big)^2 + 59238\,\frac{2}{\beta} - 41433}{18(\alpha^2 - 1)^9} \nonumber \\
 & &  + \frac{-985\big(\frac{2}{\beta}\big)^4 + 8724\big(\frac{2}{\beta}\big)^3 - 31389\big(\frac{2}{\beta}\big)^2 + 54786\,\frac{2}{\beta} - 39273}{4(\alpha^2 - 1)^8} \nonumber \\
 & & + \frac{-767\big(\frac{2}{\beta}\big)^4 + 6871\big(\frac{2}{\beta}\big)^3 - 25157\big(\frac{2}{\beta}\big)^2 + 45003\,\frac{2}{\beta} - 33321}{2(\alpha^2 - 1)^7} \nonumber \\
 & & + \frac{-3443\big(\frac{2}{\beta}\big)^4 + 31476\big(\frac{2}{\beta}\big)^3 - 118455\big(\frac{2}{\beta}\big)^2 + 219640\,\frac{2}{\beta} - 170091}{12(\alpha^2 - 1)^6} \nonumber \\
 & & + \frac{-1014\big(\frac{2}{\beta}\big)^4 + 9660\big(\frac{2}{\beta}\big)^3 - 38180\big(\frac{2}{\beta}\big)^2 + 75015\,\frac{2}{\beta} - 62150}{10(\alpha^2 - 1)^5} \nonumber \\
 & &  + \frac{-105\big(\frac{2}{\beta}\big)^4 + 1120\big(\frac{2}{\beta}\big)^3 - 4953\big(\frac{2}{\beta}\big)^2 + 10902\,\frac{2}{\beta} - 10153}{8(\alpha^2 - 1)^4} \nonumber \\
 & & + \frac{15\big(\frac{2}{\beta}\big)^3 - 128\big(\frac{2}{\beta}\big)^2 + 412\,\frac{2}{\beta} - 506}{6(\alpha^2 - 1)^3}
\eea
\end{footnotesize}

\subsection{Scaling regime and Tracy-Widom laws}

As explained in \S~\ref{subsec:TWfromLD},  we expect that if we 
 plug naively the scaling $a = \sqrt{t}(2 + N^{-2/3}\,s)$ in the large deviation of the maximal eigenvalues (Proposition~\ref{expo})
(where it is not valid as the large deviations describe fluctuations of order one from the mean value $a^*$), we should recover the right tail of the Tracy-Widom laws (that describe small fluctuations of order $N^{-2/3}$ close to $a^*$).

We first need to find the behavior of $\omega_1^{[k]}(\alpha)$ at leading order when $\alpha \rightarrow 1$ (i.e. $a \rightarrow a^* = 2\sqrt{t}$). The cases $k = 0$ and $k = 1$ can be computed from the expressions in \S~\ref{sub1}, and in general we have:
\begin{lemma}
\label{lemp}For any $k \geq 2$, there exists a polynomial $\breve{R}_{k - 1}$ of degree $k$, with integer coefficients, and a positive integer $p_{k - 1}$, such that, when $\alpha \rightarrow 1$:
\beq
\omega_1^{[k]}(\alpha) = \frac{-\breve{R}_{k - 1}\big(\frac{2}{\beta}\big)}{2^{p_{k - 1} + 1}\,(\alpha - 1)^{3k - 2}} + O\Big(\frac{1}{(\alpha - 1)^{3k - 1}}\Big)
\eeq
\end{lemma}
The proof of this technical lemma is given in \S~\ref{proof}. Since $\omega_1^{[k]}$ can be computed recursively, so do the polynomials $\breve{R}_m$. The first few can be read from \S~\ref{sub1}:
\bea
p_1 = 3 & \quad & \breve{R}_1(X) = -5X^2 + 27X - 39  \nonumber \\
p_2 = 5 & \quad & \breve{R}_2(X) = 3(-10X^3 + 73X^2 - 191X + 180) \nonumber \\
p_3 = 9 & \quad & \breve{R}_3(X) = -1105X^4 + 9720X^3 - 34557X^2 + 59238X - 41433\nonumber
\eea
If we write $a = \sqrt{t}(2 + N^{-2/3}\,s)$, we have $(\alpha - 1) \sim \sqrt{s}$ when $N$ is large, and combining the former lemma in Proposition~\ref{expo}, we find:

\begin{proposition} (Heuristics)
The probability density of the Tracy-Widom law has the following asymptotic expansion when $s \rightarrow +\infty$:
\bea
\mathsf{TW}'(s) & = & \lim_{N \rightarrow +\infty} \sqrt{t}\,N^{-2/3}\,\mathcal{G}_{N,\beta}^{\mathrm{G}\beta\mathrm{E}}\big(a = \sqrt{t}(2 + N^{-2/3}\,s)\big) \nonumber \\
& = & \frac{\Gamma\big(1 + \frac{\beta}{2}\big)}{(4\beta)^{\frac{\beta}{2}}\,\pi}\,\exp\Big\{-\frac{2\beta s^{3/2}}{3} + \Big(\frac{1}{2} - \frac{3\beta}{4}\Big)\ln s + \sum_{m \geq 1} \frac{\frac{\beta}{2}\,\breve{R}_{m}\big(\frac{2}{\beta}\big)\,s^{-\frac{3m}{2}}}{2^{p_m}\cdot 3\cdot m} \Big\} \nonumber
\eea
\end{proposition}
\noindent The polynomials $R_m$ in the result announced in Proposition~\ref{prp} are related to $\breve{R}_m$ by the triangular system:
\beq
\frac{\breve{R}_m\big(\frac{2}{\beta}\big)}{2^{p_m}\cdot 3 \cdot m} = R_m\big(\frac{2}{\beta}\big) - \sum_{\substack{m_i \geq 1\,\, (i \geq 1) \\ \sum_{i \geq 1} i\,m_i = m}} \frac{2}{\beta} \,\frac{\big(- 1 + \sum_{i \geq 1} m_i\big)!}{\prod_{i \geq 1} m_i!}\,\big(-\frac{3}{4}\big)^{\sum_{i \geq 1} m_i}\,\prod_{i \geq 1} \big[(i - 1)R_{i - 1}\big(\frac{2}{\beta}\big)\big]^{m_i} \nonumber
\eeq
with the convention $(i - 1)R_i\big|_{i = 1} \equiv 1$. We have checked that this proposition gives the correct results for the orders written above when $\beta = 1,2$ and $4$.

\subsection{Proof of Lemma~\ref{lemp}}
\label{proof}

After successive partial fraction expansions, $\omega_n^{[k]}(z_1,\ldots,z_n)$ can always be decomposed in a unique way as a sum:
\beq
\label{decom}\omega_n^{[k]}(z_1,\ldots,z_n) = \sum_r  \frac{M_{r}(\beta ; \alpha ; z_1,\ldots,z_n)}{(\alpha - 1)^{j_i(r)}\prod_{i = 1}^n (z_i - 1)^{k_i(r)}\,(z_i - \alpha^{-1})^{l_i(r)} \prod_{1 \leq i < j \leq n} (z_iz_j - 1)^{m_{i,j}(r)}}
\eeq
where $M_r(\beta ; \alpha ; z_1,\ldots,z_n)$ is a rational function of its arguments which does not vanish when one of the monomial in the denominator vanish. It may have poles when $z_i \rightarrow -1$ or $0$, when $\alpha \rightarrow 1$, but we are not interested in those poles. We define the degree of each term as $\delta(r) = \sum_{i = 1}^n j_i(r) + k_i(r) + l_i(r) + \sum_{1 \leq i < j \leq n} m_{i,j}(r)$, and we define the degree of $\omega_n^{[k]}$ as $d_{n,k} = \mathrm{max}_r\,\delta(r)$. The interest in this notion of degree is that 
$\omega_{n}^{[k]}(\alpha,\ldots,\alpha)$ must be a $O\big((\alpha - 1)^{-d_{n,k}}\big)$ when $\alpha \rightarrow 1$, and in particular:
\beq
\omega_1^{[k]}(\alpha) \in O\big((\alpha - 1)^{-d_{1,k}}\big))
\eeq

We also want to trace back the dependence of the terms of higher degree in $\beta$. Eqns.~\ref{eq1} and \ref{eq2} imply that $\omega_n^{[k]}$ is a polynomial in the variable $\frac{2}{\beta}$. Let $b_{n,k}$, the maximum power of $\frac{2}{\beta}$ appearing $M_r(\beta,\alpha ; z_1,\ldots,z_n)$ when the degree $\delta(r)$ is maximal, i.e. is equal to $d_{n,k}$.

We now determine bounds on $d_{n,k}$ and $b_{n,k}$, by recursion on $n + k$. Lemma~\ref{lemp} is then an application of the lemma below to $n = 1$.
\begin{lemma}
For all $(n,k) \neq (1,0)$:
\bea
d_{n,k} \leq 4n + 3k - 6,\qquad b_{n,k} \leq n + k - 1 
\eea
\end{lemma}

\noindent \textbf{Proof.} When $n + k = 2$, we already know the answer from Eqn.~\ref{eq11} and \ref{eq20}, namely:
\bea
d_{1,1} = 1 \quad & & \quad d_{2,0} = 2  \\
b_{1,1} = 1 \quad & & \quad b_{2,0} = 1
\eea
and it satisfies the desired property. Let $p \geq 2$ and assume the lemma whenever $n + k \leq p$. When $n + k = p + 1$, Eqn.~\ref{eq1} or Eqn.~\ref{eq2} computes $\omega_{n}^{[k]}$ in terms of $\omega_{n'}^{[k']}$ with $n' + k' \leq p$. This relation can be organized as follows:
\bea
\omega_{n}^{[k]}(z,z_I) & = & \frac{-1}{Y(x(z))\,x'(z)}\Big[A_{n + 1}^{[k - 2]}(z ; z_I) + B_{n + 1}^{[k]}(z ; z_I) + \Big(1 - \frac{2}{\beta}\Big) C_{n}^{[k - 1]}(z ; z_I) \nonumber \\
\label{pro} & & +  2\,D_n^{[k - 1]}(z ; z_I) + \frac{2}{\beta}\,\sum_{i \in I} E_{n - 1}^{[k]}(z,z_i ; z_{I \setminus \{i\}}) \Big]
\eea
where:
\bea
A_{n + 1}^{[k + 2]}(z ; z_I) & = & \omega_{n + 1}^{[k + 2]}(z,z,z_I) \\
B_{n + 1}^{[k]}(z ; z_I) & = &  \sum_{\substack{J \subseteq I\quad 0 \leq k' \leq k \\ (J,k') \neq (\emptyset,0),(I,k')}} \omega_{|J| + 1}^{[k]}(x,x_J)\,\omega_{n - |J|}^{[k - k']}(x,x_{I\setminus J}) \\
C_{n}^{[k - 1]}(z ; z_I) & = & x'(z)\,\partial_z\Big(\frac{\omega_n^{[k - 1]}(z,z_I)}{x'(z)}\Big)  \\
D_n^{[k]}(z ; z_I) & = & \big(x'(z)\big)^2\,z\,\frac{\frac{\omega_{n}^{[k - 1]}(z,z_I)}{x'(z)} - \frac{\omega_n^{[k - 1]}(\alpha,z_I)}{x'(\alpha)}}{(z - \alpha)(z - \alpha^{-1})}  \\
E_{n - 1}^{[k]}(z,z_i ; z_J) & = & \big(x'(z)\big)^2\,\partial_{z_i}\Big(z\,z_i\,\frac{\frac{\omega_{n - 1}^{[k]}(z,z_J)}{x'(z)} - \frac{\omega_{n - 1}^{[k]}(z_i,z_J)}{x'(z_i)}}{(z - z_i)(zz_i - 1)}\Big) 
\eea
The very definition of the degree also allow to follow the degree of each of the terms:
\begin{itemize}
\item[$\diamond$] The degree of $A_{n + 1}^{[k + 2]}$ is atmost $d_{n + 1,k - 2}$, which is by assumption smaller than $4n + 3k - 8$.
\item[$\diamond$] The degree of $B_{n + 1}^{[k]}$ is bounded by the maximum of $(d_{n' + 1,k'} + d_{n - n',k - k'})$ for $0 \leq n' \leq n$, $0 \leq k' \leq k$ and $(n',k') \neq (0,0),(n,k)$. But since $d_{n',k'}$ is bounded by an affine function of $n'$ and $k'$ for $n' + k' < n + k$ by assumption, all these numbers are bounded by the same quantity, namely $(4(n' + 1) + 3k' - 6) + (4(n - n') + 3(k - k')  - 6) = 4n + 3k - 8$.
\item[$\diamond$] The degree of $C_{n}^{[k - 1]}$ is $d_{n,k - 1} + 1 \leq 4n + 3k - 8$.
\item[$\diamond$] A term in $\omega_{n}^{[k- 1]}$ containing a factor $(z - \alpha^{-1})^{-l}$, yield a term containing a factor $(z - \alpha^{-1})^{-(l + 1)}$ in $D_{n}^{[k - 1]}$. Thus, the degree of $D_{n}^{[k - 1]}$ is bounded by $d_{n,k - 1} + 1 \leq 4n + 3k - 8$.
\item[$\diamond$] A term in $\omega_{n - 1}^{[k]}$ containing a factor $(z_1 - 1)^{-j}$ yield in the ratio $E_{n - 1}^{[k]}(z,z_i ; z_I)$ several terms, with possible factors $(z - 1)^2(z_i - 1)^{-(j + 3)}(zz_i - 1)^{-1}$, $(z - 1)^2(z_i - 1)^{-(j + 2)}(zz_i - 1)$, $(z - 1)^{2 - (j + 2)}(zz_i - 1)^{-2}$ or $(z - 1)^{2 - (j + 3)}(zz_i - 1)^{-1}$. Thus, the degree of $D_{n - 1,k}$ is bounded by $d_{n - 1,k} + 2 = 4n + 3k - 8$.
\end{itemize}
Besides, the quantity $Y(x(z))\,x'(z) = -z^{-3}(z - 1)^2(z + 1)^2$ has degree $2$. So, we conclude that for $n + k = p + 1$, we have $d_{n,k} \leq (4n + 3k - 8) + 2 = 4n + 3k - 6$ which is the desired bound.

Let us come to the dependence in $\beta$. We know that $Y(x(z))x'(z)$ does not depend on $\beta$. Eqn.~\ref{pro} implies that $b_{n,k}$ is bounded by the maximum among the numbers $b_{n + 1,k - 2}$, $b_{n' + 1,k'} + b_{n - n',k - k'}$ for $0 \leq n' \leq n$, $0 \leq k' \leq k$ and $(n',k') \neq (0,0),(n,k)$, $(b_{n,k - 1} + 1)$ and $(b_{n - 1,k} + 1)$. And the recursion hypothesis implies that all those numbers are bounded by $n + k - 1$, hence the bound for $b_{n,k}$. \hfill $\Box$

\vspace{0.2cm}

\noindent Eqn.~\ref{pro} implies that $\omega_{n}^{[k]}(z_1,\ldots,z_n)$ is a rational fraction in $z_1,\ldots,z_n$, and also of the variable $\alpha$, with rational coefficients. We would like to know which denominators do we get, especially in front of the leading term when $\alpha \rightarrow 1$. Notice that Eqn.~\ref{pro} only involves integer coefficients, and the only poles in the partial fraction expansion of $M_r(\beta,\alpha ; z_1,\ldots,z_n)$ (cf. Eqn.~\ref{decom}) are of the form $(\alpha + 1)^{-j'_i(r)}\prod_{i = 1}^n (z_i + 1)^{-k_i'(r)}$. So, we can only get a power of $2$ in the denominator, coming from expansion of $(\alpha + 1)^{-j_i(r)'}\prod_i (z + 1)^{-k_i'(r)}$ when $z_i \rightarrow \alpha$ and  $\alpha \rightarrow 1$. This justifies the factorization of a power of $2$ (which we do not try to compute here) in Lemma~\ref{lemp}, and the fact that $\breve{R}_m$ is a polynomial with integer coefficients.

\subsection*{Addendum}

While we were completing this work, P.~Forrester \cite{Forr} has studied the two first leading terms and  the constant prefactor in the large deviation function, and its heuristic matching with the right tail of Tracy-Widom $\beta$ laws, with similar results. His method relies on the functional equation satisfied by the spectral density and its first correction. This corresponds in this article to the loop equation \ref{28} up to $O(1)$, in which case the term $W_2(x,x)$ can be forgotten and the equation only involve the spectral density itself. \cite{Forr} also obtain the two first orders and the constant prefactor of the large deviation function in the Laguerre ensemble. As we pointed out in \S~\ref{subsec:corrgauss}, the method of loop equations could be applied to obtain recursively, without any inversion of linear operator, the full asymptotic expansion of the large deviation function.

\subsection*{Acknowledgments}

We thank F.~David, B.~Eynard, C.~Hagendorf, S.N.~Majumdar for fruitful discussions, and A.~Comtet for encouragements to complete this project. The work of G.B. benefited from the ANR project Grandes Matrices Al\'eatoires ANR-08-BLAN-0311-01.

\bibliographystyle{amsalpha}
\bibliography{BibliTW2}

\providecommand{\bysame}{\leavevmode\hbox to3em{\hrulefill}\thinspace}
\providecommand{\MR}{\relax\ifhmode\unskip\space\fi MR }
\providecommand{\MRhref}[2]{%
  \href{http://www.ams.org/mathscinet-getitem?mr=#1}{#2}
}
\providecommand{\href}[2]{#2}
\begin{thebibliography}{BEMN11}

\bibitem[AD99]{AD}
D.~Aldous and P.~Diaconis, \emph{{Longest increasing subsequences: from
  patience sorting to the Baik-Deift-Johansson theorem}}, Bull. Amer. Math.
  Soc. \textbf{36} (1999), 413--432.

\bibitem[BBd08]{BBdiF}
J.~Baik, R.~Buckingham, and J.~di{F}ranco, \emph{Asymptotics of {T}racy-{W}idom
  distributions and the total integral of a {P}ainlev\'{e} {II} function},
  Commun. Math. Phys. \textbf{280} (2008), no.~2, 463--497,
  \href{http://arxiv.org/abs/0704.3636}{\textsf{math.FA/0704.3636}}.

\bibitem[BEMN11]{BEMN}
G.~Borot, B.~Eynard, S.N. Majumdar, and C.~Nadal, \emph{Large deviations of the
  maximal eigenvalue of random matrices}, to appear in J. Stat. Phys. (2011),
  \href{http://arxiv.org/abs/1009.1945}{\textsf{math-ph/1009.1945}}.

\bibitem[BG11]{BG11}
G.~Borot and A.~Guionnet, \emph{Asymptotic expansion of $\beta$ matrix models
  in the one-cut regime},
  \href{http://arxiv.org/abs/1107.1167}{\textsf{math-PR/1107.1167}}.

\bibitem[BV10]{VirBloe}
A.~Bloemendal and B.~Vir{\'{a}}g, \emph{Limits of spiked random matrices {I}},
  \href{http://arxiv.org/abs/1011.1877}{\textsf{math.PR/1011.1877}}.

\bibitem[CE06]{CE06}
L.O. Chekhov and B.~Eynard, \emph{Matrix eigenvalue model: {F}eynman graph
  technique for all genera}, JHEP (2006), no.~0612:026,
  \href{http://arxiv.org/abs/math-ph/0604014}{\textsf{math-ph/0604014}}.

\bibitem[DE02]{DE02}
I.~Dumitriu and A.~Edelman, \emph{Matrix models for beta ensembles}, J. Math.
  Phys. \textbf{43} (2002), no.~11, 5830--5847,
  \href{http://arxiv.org/abs/math-ph/0206043}{\textsf{math-ph/0206043}}.

\bibitem[Dei07]{Deiftuni}
P.~Deift, \emph{Universality for mathematical and physical systems}, Proceeding
  of the ICM (2007), 125--152.

\bibitem[DG07]{DG07}
P.~Deift and D.~Gioev, \emph{Universality at the edge of the spectrum for
  unitary, orthogonal and symplectic ensembles of random matrices}, Comm. Pure
  Appl. Math. \textbf{60} (2007), no.~6, 867--910,
  \href{http://arxiv.org/abs/math-ph/0507023}{\textsf{math-ph/0507023}}.

\bibitem[DIK08]{DIK}
P.~Deift, A.~Its, and I.~Krasovsky, \emph{Asymptotics of the {A}iry-kernel
  determinant}, Commun. Math. Phys. \textbf{278} (2008), no.~3, 643--678,
  \href{http://arxiv.org/abs/math.FA/0609451}{\textsf{math.FA/0609451}}.

\bibitem[DV11]{VirDum}
L.~Dumaz and B.~Vir{\'{a}}g, \emph{The right tail exponent of the
  {T}racy-{W}idom-beta distribution},
  \href{http://arxiv.org/abs/1102.4818}{\textsf{math.PR/1102.4818}}.

\bibitem[Dys62]{Dyson}
F.~Dyson, \emph{Statistical theory of the energy levels of complex systems}, J.
  Math. Phys. \textbf{3} (1962), 140.

\bibitem[FN80]{FN80}
H.~Flaschka and A.C. Newell, \emph{Monodromy- and spectrum-preserving
  deformations, {I}}, Commun. Math. Phys. \textbf{76} (1980), no.~1, 65--116.

\bibitem[For10]{forrester}
P.~J. Forrester, \emph{Log-gases and random matrices}, Princeton University
  Press, Princeton, New Jersey, 2010.

\bibitem[For11]{Forr}
P.J. Forrester, \emph{Spectral density asymptotics for {G}aussian and
  {L}aguerre $\beta$-ensembles in the exponentially small region},
  \href{http://arxiv.org/abs/1111.1350}{\textsf{math-ph/1111.1350}}.

\bibitem[FT28]{FT28}
R.A. Fisher and L.H.C. Tippett, \emph{Limiting forms of the frequency
  distribution of the largest or smallest member of a sample}, Proceedings of
  the {C}ambridge {P}hilosophical {S}ociety \textbf{24} (1928), 189--190.

\bibitem[FW01]{FW01}
P.J. Forrester and N.S. Witte, \emph{Application of the $\tau$-function theory
  of {P}ainlev\'{e} equations to random matrices}, Commun. Math. Phys.
  \textbf{219} (2001), 357--398,
  \href{http://arxiv.org/abs/math-ph/0103025}{\textsf{math-ph/0103025}}.

\bibitem[Gne43]{Gne43}
B.V. Gnedenko, \emph{Sur la distribution limite du terme maximum d'une
  s\'{e}rie al\'{e}atoire}, Ann. Math. \textbf{44} (1943), 423--453.

\bibitem[Gum58]{Gum58}
E~Gumbel, \emph{Statistics of extremes}, Columbia University Press, NY, 1958.

\bibitem[HM80]{HMcL}
S.P. Hastings and J.B. Mc{L}eod, \emph{A boundary value problem associated with
  the second {P}ainlev\'{e} transcendent and the {K}orteweg-de {V}ries
  equation}, Archive for {R}ational {M}echanics and {A}nalysis \textbf{73}
  (1980), no.~1, 31--51.

\bibitem[Joh98]{Johan}
K.~Johansson, \emph{On fluctuations of eigenvalues of random hermitian
  matrices}, Duke Math. J. \textbf{91} (1998), 151--204.

\bibitem[KK10]{krug}
T.~Kriecherbauer and J.~Krug, \emph{{A pedestrian's view on interacting
  particle systems, KPZ universality and random matrices}}, J. Phys. A.: Math.
  Theor. \textbf{43} (2010), 403001,
  \href{http://arxiv.org/abs/0803.2796}{\textsf{cond.mat-stat.mech/0803.2796}}.

\bibitem[Maj07]{maj-review}
S.~N. Majumdar, \emph{{Complex Systems (Les Houches lecture notes)}},
  ch.~{Random matrices, the Ulam problem, directed polymers and growth models,
  and sequence matching}, pp.~179--216, Elsevier, Amsterdam, 2007,
  \href{http://arxiv.org/abs/cond-mat/0701193}{\textsf{cond.mat-stat.mech/0701193}}.

\bibitem[Meh04]{MehtaBook}
M.L. Mehta, \emph{Random matrices}, 3$^{\textrm{\`{e}me}}$ ed., Pure and
  {A}pplied {M}athematics, vol. 142, Elsevier/Academic, Amsterdam, 2004.

\bibitem[MV09]{MajVer}
S.N. Majumdar and M.~Vergassola, \emph{Large deviations of the maximum
  eigenvalue for {W}ishart and {G}aussian random matrices}, Phys. Rev. Lett
  \textbf{102} (2009), no.~060601,
  \href{http://arxiv.org/abs/0811.2290}{\textsf{cond-mat.stat-mech/0811.2290}}.

\bibitem[NM11]{NadMaj}
C.~Nadal and S.~Majumdar, \emph{A simple derivation of the {T}racy-{W}idom
  distribution of the maximal eigenvalue of a {G}aussian unitary random
  matrix}, J. Stat. Phys. (2011), no.~P04001,
  \href{http://arxiv.org/abs/1102.0738}{\textsf{cond-mat.stat-mech/1102.0738}}.

\bibitem[PS00]{Sph}
M.~Pr\"{a}hofer and H.~Spohn, \emph{Universal distributions for growth
  processes in 1+1 dimensions and random matrices}, Phys. Rev. Lett.
  \textbf{84} (2000), no.~21, 4882--4885.

\bibitem[RRV06]{RRV}
J.A. Ram{\`i}rez, B.~Rider, and B.~Vir{\'a}g, \emph{Beta ensembles, stochastic
  {A}iry process, and a diffusion},
  \href{http://arxiv.org/abs/math.FA/0607331}{\textsf{math.FA/0609451}}.

\bibitem[TS10]{Takeuchi}
K.~Takeuchi and M.~Sano, \emph{Universal fluctuations of growing interfaces:
  evidence in turbulent liquid crystals}, Phys. Rev. Lett. \textbf{104} (2010),
  no.~230601,
  \href{http://arxiv.org/abs/1001.5121}{\textsf{cond-mat.stat-mech/1001.5121}}.

\bibitem[TW94]{TW92}
C.A. Tracy and H.~Widom, \emph{Level spacing distributions and the {A}iry
  kernel}, Commun. Math. Phys. \textbf{159} (1994), 151--174,
  \href{http://arxiv.org/abs/hep-th/9211141}{\textsf{hep-th/9211141}}.

\bibitem[TW96]{TW95}
\bysame, \emph{On orthogonal and symplectic matrix ensembles}, Commun. Math.
  Phys. \textbf{177} (1996), 727--754,
  \href{http://arxiv.org/abs/solv-int/9509007}{\textsf{solv-int/9509007}}.

\bibitem[TW02]{TracyReview}
\bysame, \emph{Distribution functions for largest eigenvalues and their
  applications}, Proceedings of the ICM \textbf{1} (2002), 587--596,
  \href{http://arxiv.org/abs/math-ph/0210034}{\textsf{math-ph/0210034}}.

\bibitem[Wig58]{Wig58}
E.P. Wigner, \emph{On the distribution of the roots of certain symmetric
  matrices}, Ann. of Math. (2) \textbf{67} (1958), 325--327.

\end{thebibliography}

\end{document}